# Drug discovery guided by maximum drug likeness

Hao-Yu Zhu[1], Shi-Jie Du[1], Lu Xu[1,2]*, Wei Shi[1]*

[1] College of Material and Chemical Engineering, Tongren University, Tongren 554300, PR China

[2] School of Sports and Health Science, Tongren University, Tongren, 554300, PR China

**Abstract** To overcome the high attrition rate and limited clinical translatability in drug discovery, we introduce the concept of Maximum Drug-Likeness (MDL) and develop an applicable Fivefold MDL strategy (5F-MDL) to reshape the screening paradigm. The 5F-MDL strategy integrates an ensemble of 33 deep learning sub-models to construct a 33-dimensional property spectrum that quantifies the global phenotypic alignment of candidate molecules with clinically approved drugs along five axes: physicochemical properties, pharmacokinetics, efficacy, safety, and stability. Using drug-likeness scores derived from this 33-dimensional profile, we prioritized 15 high-potential molecules from a 16-million-molecule library. Experimental validation demonstrated that the lead compound M2 not only exhibits potent antibacterial activity, with a minimum inhibitory concentration (MIC) of 25.6 μg/mL, but also achieves binding stability superior to cefuroxime, as indicated by Molecular Mechanics Poisson-Boltzmann surface area (MM-PBSA) calculations of -38.54 kcal/mol and a root-mean-square deviation (RMSD) of 2.8 Å. This strategy could overcome scaffold constraints and offers an efficient route for discovering lead compounds with favorable prospects against drug-resistant bacteria.

Corresponding authors：

Email: lxchemo@163.com (Xu L.); chysw@gztrc.edu.cn (Shi W.)

# 1. Introduction

Drug discovery remains central to the fight against major diseases. For high-burden conditions such as cancer, human immunodeficiency virus/acquired immunodeficiency syndrome (HIV/AIDS), and tuberculosis, each advance in pharmaceutical research and development can improve treatment for tens of millions of patients [1-6]. Nevertheless, the enterprise is marked by persistent inefficiency, protracted timelines, and high attrition. Of the tens of thousands of synthesized and tested compounds, approximately 100 are advanced for development consideration, about 10 enter clinical trials, and only 1 is ultimately approved; the entire process often exceeds a decade [7-9]. The tension between urgent medical need and high failure rates is driving the emergence and refinement of more efficient and systematic discovery paradigms.

Drug research and development has progressed from empirical practice to systematic, rational design. Early trial-and-error work in lead optimization and activity screening laid the foundation [10-13]. Later, high-throughput screening (HTS) and rational drug design (RDD) became mainstream. HTS enables rapid testing of very large libraries [14,15], but it often yields many false positives, shows limited clinical translatability, and entails high costs and technical barriers [16-20]. RDD uses three-dimensional target structures and mechanism-of-action models to improve specificity and efficiency [21-23], but it depends on reliable structural data and well-understood mechanisms [24-26]. Computer-aided drug design (CADD) integrates molecular docking, quantitative structure-activity relationship (QSAR) modeling, molecular dynamics (MD) simulations, and pharmacophore methods [27,28]. In this toolkit, molecular docking predicts binding modes and affinities [29,30]; QSAR builds quantitative models of structure-activity relationships [31,32]; MD probes interaction stability and allosteric effects [33,34]; and pharmacophore modeling captures three-dimensional activity features and guides virtual screening [35-37]. These methods have also been valuable in public health emergencies [38,39]. However, current CADD practice has long focused on efficacy (activity and affinity), with limited attention to key dimensions of druggability, including physicochemical developability, absorption, distribution, metabolism, and excretion (ADME), safety, and molecular stability. This imbalance has contributed to failures during the transition from lead to candidate and to rising development costs [16-20,24-26,29-37]. Therefore, a systematic framework that centers on multidimensional developability, while balancing efficiency and translatability, is highly required.

In response to these gaps, we propose Maximum Drug-Likeness (MDL) and formulate the Fivefold MDL (5F-MDL) paradigm. In this paradigm, the screening objective is integrated similarity to approved drugs across five dimensions: physicochemical properties, pharmacokinetics, efficacy, safety, and stability. Candidate molecules must achieve high similarity in all five dimensions. To realize this objective, we use large-scale molecular structure and activity data to build, within an ensemble deep learning framework, a set of quantitative structure-activity relationship (QSAR) models that reliably predict properties across the five dimensions. We then prioritize high-potential molecules that meet the 5F-MDL criteria from high-throughput screening libraries. We further use experimental validation and molecular docking to calibrate predictions and provide mechanistic support [27-34]. This approach aims to reduce downstream failures caused by single-dimension optimization and to improve overall developability and success rates.

Within this framework, we adopted Escherichia coli (E. coli) as the reference pathogen for molecular selection and evaluated the Fivefold Maximum Drug-Likeness (5F-MDL) method under realistic, high-threat, drug-resistant conditions. Escherichia coli is a common Gram-negative opportunistic pathogen and a major public health threat [40-43]. In 2024, the World Health Organization (WHO) continued to classify carbapenem-resistant Enterobacteriaceae (CRE) as critical priority pathogens [44]. Escherichia coli is a leading cause of urinary tract infections and is associated with multiple hospital-acquired infections [45,46]. Clinical treatment has relied primarily on beta-lactam antibiotics that target penicillin-binding proteins (PBPs), with penicillin-binding protein 2 (PBP2) playing a pivotal role [47-49]. However, enzyme-mediated hydrolysis, reduced membrane permeability, target-site mutations, and bypass pathways limit drug affinity and exposure, leading to therapeutic failure [50-60]. The global burden of resistance continues to rise, with high isolation rates reported in some regions [61,62], while innovation in new scaffolds for Gram-negative pathogens remains limited [63-65]. Against this backdrop, the 5F-MDL paradigm is designed to maintain activity against key targets such as PBP2 while jointly optimizing physicochemical, pharmacokinetic, safety, and stability attributes to enhance clinical translatability [47-49,53-56].

In summary, we developed and validated a 5F-MDL strategy and an ensemble deep learning-based screening workflow for E. coli molecular screening. Molecular docking and experimental data supported the conclusions and feasibility [65]. We anticipate that this framework will serve as a novel early screening paradigm for complex drug-resistant pathogens and improve research and development efficiency and success in practice.

## 2. Materials and Methods

### 2.1 Maximum Drug-Likeness (MDL) and Fivefold Maximum Drug-Likeness (5F-MDL)

To increase the efficiency of drug screening, we introduce Maximum Drug-Likeness (MDL), defined as the highest attainable similarity, within the widest feasible chemical space, between candidate molecules and clinically approved drugs. Based on this concept, we develop an operational implementation, Fivefold Maximum Drug-Likeness (5F-MDL). This framework evaluates five dimensions: physicochemical properties; pharmacokinetics (absorption, distribution, metabolism, and excretion, ADME); efficacy; safety; and stability (Figure 1). Under the 5F-MDL paradigm, candidate molecules must achieve maximal integrated similarity to approved drugs across all five dimensions.

[Please insert Fig. 1 here.]

Physicochemical properties are the fundamental physical and chemical attributes of a molecule and define its developability window. We consider molecular weight; calculated partition coefficient (cLogP); distribution coefficient (logD); acid dissociation constant (pKa); hydrogen bond donor and acceptor counts; topological polar surface area (tPSA); and aqueous solubility. These values are expected to fall within distributions typical of approved drugs to maintain solubility, permeability, and formulation feasibility.

Pharmacokinetics describes whole-body disposition, namely absorption, distribution, metabolism, and excretion (ADME). It includes absorption and bioavailability; distribution (volume of distribution, Vd; tissue penetration; and plasma protein binding); metabolic pathways and clearance; and terminal half-life.

Efficacy is the true pharmacological effect in relevant models and its linkage to the intended target. Emphasis is placed on potency metrics (half-maximal inhibitory concentration, IC50; half-maximal effective concentration, EC50; and minimum inhibitory concentration, MIC), target occupancy and confirmation of the mechanism of action (MoA), pharmacodynamic exposure-response relationships (PD), and the effective range for the intended indication, so that the desired therapeutic effect is achieved at clinically attainable exposures.

Safety covers tolerability and risk boundaries. It includes cellular and organ toxicities, genotoxicity, cardiovascular liabilities (for example, inhibition of the human ether-a-go-go-related gene channel, hERG), hepatic and renal risks, immunological and endocrine effects, drug-drug interaction liability, and therapeutic index. The objective is to achieve risk profiles comparable to or better than those of approved drugs.

Stability refers to the preservation of structure and activity from research through clinical use. It includes chemical and physical stability (thermal, photolytic, oxidative, and hydrolytic), solid-state and solution stability, plasma and metabolic stability, and compatibility with shelf-life, transport, and storage requirements, thereby ensuring quality control and clinical usability.

Given the availability of quantitative structure-activity relationship (QSAR) data, 33 specific properties spanning these five dimensions were selected as feature dimensions (see Table 1).

[Please insert Table 1 here.]

**2.2 Data**

### 2.2.1 Clinically Approved Anti-Escherichia coli Drug Set

This dataset includes three clinically approved drugs with activity against Escherichia coli (E. coli). The agents are cephalosporin antibiotics within the beta-lactam class. Their principal molecular target is penicillin-binding protein 2 (PBP2), a key enzyme in bacterial cell wall biosynthesis. These agents are important options for treating infections caused by susceptible E. coli, with primary indications that include urinary tract infections, intra-abdominal infections, bacteremia, and pelvic inflammatory disease. The names and chemical structures of the three drugs are shown in Figure 2.

[Please insert Fig. 2 here.]

### 2.2.2 Candidate Molecule Set

Screening libraries were obtained from eMolecules (San Diego, California, United States), TargetMol Chemicals Inc. (Shanghai, China), and MedChemExpress (Shanghai, China). In total, approximately 16 million commercially available small molecules with diverse chemical structures and well-characterized physicochemical properties were included in the screening set. Most compounds conformed to Lipinski's Rule of Five.

### 2.3 Drug Screening Guided by Fivefold Maximum Drug-Likeness

To enable screening guided by 5F-MDL, a QSAR framework based on an ensemble of deep neural networks was developed. The framework was organized into three modules (see Figure 3): (i) supervised pretraining of ensemble deep learning models on 33 molecular properties; (ii) maximum-similarity (Max-Sim) transfer scoring relative to reference drugs; and (iii) generation of the final drug-likeness score, $S_{5F}$. A transfer learning paradigm was adopted. Design details, mathematical formulations, training protocols, and robustness control measures are provided below.

[Please insert Fig. 3 here.]

### 2.3.1 Molecular Input and Representation

Each molecule was represented by 3,599 two-dimensional molecular descriptors computed with Dragon 7. These descriptors span atom-level, bond-level, and whole-molecule features. They were used to build quantitative structure-activity relationship (QSAR) models for 33 properties across five domains: physicochemical properties, pharmacokinetics, efficacy, safety, and stability (see Table 1). Together, these 33 properties provide a multidimensional view of molecular characteristics from five complementary perspectives, with the goal of closely approximating the MDL feature space relevant to biological effects and environmental safety.

**2.3.2 Supervised Pretraining of the Deep Ensemble**

Independent predictor submodels were trained for each of the 33 key molecular properties (Table 1). Training data were compiled from publicly labeled sources, including ChEMBL, PubChem, ToxCast, and the United States Environmental Protection Agency Distributed Structure-Searchable Toxicity (EPA DSSTox) database, as well as the QSAR literature. To promote generalization, the downstream screening libraries (eMolecules, TargetMol, and MedChemExpress) were not included in their entirety in the training set, although limited overlap at the molecule level could not be excluded. For each property, data were randomly partitioned into training, validation, and test sets in approximately a 70%:15%:15% ratio. Feature selection was performed using the scikit-learn feature_selection module (Python 3.6) with random forest models. Feature importances were estimated, and features exceeding a predefined importance threshold were retained via the SelectFromModel procedure, in order to improve the performance of the subsequent deep learning models on both regression and classification tasks.

All 33 submodels adopted a deep neural network architecture (Figure 4) to extract robust property representations from high-dimensional molecular features. The network comprised three fully connected hidden layers of sizes 2,048, 1,024, and 128. All hidden layers used the rectified linear unit (ReLU) activation function and were regularized with dropout (rate = 0.2) and weight decay ($1 \times 10^{-2}$) to mitigate overfitting. The third hidden layer (128 units) served as a bottleneck, enforcing a compact and discriminative latent molecular representation. The output layer depended on task type: for regression tasks, a single linear node was used to predict continuous values; for classification tasks, a single node with sigmoid activation was employed to produce the probability of the positive class. Models were trained using the AdamW optimizer (learning rate = $1 \times 10^{-4}$; batch size = 128) with early stopping (patience = 10).

For regression tasks, the mean squared error (MSE) loss was used:

$$\text{MSE} = \frac{1}{N_{batch}} \sum_{i=1}^{N_{batch}} (y_i - \hat{y}_i)^2 \tag{1}$$

where $N_{batch}$ denotes the batch size, $y_i$ is the reference value, and $\hat{y}_i$ is the prediction, respectively.

For classification tasks, the binary cross-entropy loss was used:

$$L = -\frac{1}{N_{batch}} \sum_{i=1}^{N_{batch}} [y_i \log(\hat{y}_i) + (1 - y_i) \log(1 - \hat{y}_i)] \tag{2}$$

where $y_i \in \{0,1\}$ denotes the reference label and $\hat{y}_i \in [0,1]$ denotes the predicted probability.

Evaluation on the test sets used the coefficient of determination ($R^2$) for regression tasks and the area under the receiver operating characteristic curve (ROC AUC) for classification tasks. Performance on the independent test sets is summarized in Table 2.

To quantify candidate–reference similarity in a common mathematical space, a fivefold property spectrum was generated using the 33 trained submodels. For any molecule, this spectrum was represented as a vector $V \in R^{33}$. Because the raw properties span heterogeneous physical scales (for example, boiling point on the order of 300–600 K versus toxicity probabilities in [0, 1]), direct comparisons would be dominated by large-magnitude quantities. Accordingly, normalization procedures were applied. For regression-type properties that span orders of magnitude (for example, minimum inhibitory concentration (MIC), inhibition constant (Ki), and median lethal dose (LD50)), log transformations were applied during model training. At prediction time, outputs were min-max normalized using the training-set statistics, with clipping to the unit interval to guard against out-of-distribution values:

$$v_k = \text{clip}\left(\frac{y_k - \min(Y_{train}^k)}{\max(Y_{train}^k) - \min(Y_{train}^k)}, 0, 1\right) \quad (3)$$

where $y_k$ is the raw prediction for property k, $Y_{train}^k$ is the set of training labels for property k, and $\text{clip}(x, 0, 1) = \max(0, \min(x, 1))$. This ensured that each normalized component $v_k$ lies in [0, 1]. For classification properties (for example, human Ether-à-go-go-Related Gene (hERG) risk and chemical stability), the model's positive-class probability was used directly, which naturally lies in [0, 1] and preserves confidence information.

Through these steps, each molecular property spectrum $V$ was mapped to the 33-dimensional unit hypercube $[0,1]^{33}$. Because the values in $V$ carry absolute biological meaning (that is, magnitude reflects property strength), direction-only similarity measures such as cosine similarity may underweight absolute differences. A Euclidean distance-based measure was therefore used to compute the final composite score $S_{5F}$:

$$S_{5F}(V_{cand}, V_{ref}) = 1 - \frac{\|V_{cand} - V_{ref}\|_2}{\sqrt{N_{prop}}} \quad (4)$$

where $V_{cand}$ and $V_{ref}$ are the normalized property vectors of the candidate and reference molecules, respectively, $N_{prop} = 33$ is the dimensionality, and $\|\cdot\|_2$ denotes the Euclidean norm. The denominator $\sqrt{N_{prop}}$ is the maximum theoretical distance in this unit hypercube, ensuring $S_{5F} \in [0,1]$ and approaching 1.0 only when the candidate and reference are very close on all 33 dimensions. The final score for a candidate molecule was defined as the maximum $S_{5F}$ across all reference drugs.

[Please insert Fig. 4 here.]

**2.4 External Validation and Evaluation Metrics**

On the basis of the computed composite score $S_{5F}$, a ranking-based prioritization strategy was implemented. To comprehensively cover the bioactivity profiles of distinct reference drugs, candidate molecules were separately matched against three reference drugs, ranked in descending order of score within each group, and the top five per group were retained. This stratified selection ensured close alignment with the reference profiles in the high-dimensional property space while preserving scaffold diversity. In total, fifteen top-ranked candidate small molecules were selected for experimental validation. Validation was conducted at three levels: in vitro antibacterial activity testing, structure-based molecular docking, and molecular dynamics (MD) simulations to model dynamic binding.

### 2.4.1 Antibacterial Activity Testing

An adapted Kirby–Bauer disk diffusion assay was employed for preliminary antibacterial screening of the fifteen candidates. Escherichia coli (E. coli) ATCC 25922 was inoculated into Mueller–Hinton broth (MHB), cultured at 37 °C to the logarithmic phase, and adjusted with sterile normal saline to a turbidity equivalent to a 0.5 McFarland standard (approximately $1.5 \times 10^8$ colony-forming units per milliliter, CFU/mL). The bacterial suspension was then uniformly spread onto Mueller–Hinton agar (MHA) plates.

Candidate molecules and positive control antibiotics (cefradine, cefuroxime, and ceftriaxone) were dissolved in dimethyl sulfoxide (DMSO), and drug-impregnated disks were prepared (30 μg per disk; DMSO content maintained below 1% to exclude solvent effects). Disks were placed on the inoculated plates, which were incubated inverted at 37 °C for 16–18 hours. Inhibition zone diameters (millimeters) were measured and recorded, and compounds exhibiting visible antibacterial activity were advanced to minimum inhibitory concentration (MIC) determination. All experiments were performed in triplicate as independent biological replicates.

### 2.4.2 Determination of Minimum Inhibitory Concentration (MIC) and Minimum Bactericidal Concentration (MBC)

On the basis of the disk diffusion results, active candidates were evaluated by broth microdilution in accordance with the Clinical and Laboratory Standards Institute (CLSI) guideline M07-A9 to determine the MIC. Stock solutions were prepared in dimethyl sulfoxide and serially twofold diluted in cation-adjusted Mueller–Hinton broth (CA-MHB) to yield final well concentrations spanning 1.6–102.4 μg/mL. An inoculum adjusted to $5 \times 10^5$ CFU/mL was added to compound-containing microplates, which were incubated at 37 °C for 24 hours. The MIC was defined as the lowest concentration at which no visible bacterial growth was observed by the unaided eye.

Following MIC determination, 10 μL from each well without visible growth was plated onto MHA and incubated at 37 °C for 24 hours. The minimum bactericidal concentration (MBC) was defined as the lowest concentration resulting in a reduction in colony count of ⩾99.9%. Solvent (DMSO) and growth controls were included, and all measurements were performed in three independent biological replicates.

### 2.4.3 Molecular Docking Validation

In accordance with published procedures [66], penicillin-binding protein 2 (PBP2; Protein Data Bank (PDB) ID: 6G9P) was selected as the receptor. The protein structure was downloaded from the Research Collaboratory for Structural Bioinformatics Protein Data Bank (RCSB PDB), and the co-crystallized ligand and water molecules were removed. Three-dimensional structures of the candidate molecules and the reference drug (cefuroxime) were constructed using ChemDraw Professional 15.0 and saved in Structure Data File (SDF) format.

Blind docking was conducted on the CB-Dock2 platform [67]. Five potential binding cavities were identified, with volumes ranging from 286 to 635 Å$^3$ and exhibiting distinct geometries and interaction potentials. Docking outputs (plain-text TXT files) were downloaded, and the protein–ligand complexes and interaction patterns were visualized using PyMOL 3.1.0 and BIOVIA Discovery Studio 2021.

### 2.4.4 Molecular Dynamics Simulation Validation

To assess dynamic stability and binding persistence under physiologically relevant conditions, the most active M2–PBP2 complex and the positive-control cefuroxime–PBP2 complex were subjected to molecular dynamics simulations using the AMBER 22 software suite. Protein parameters were assigned using the ff14SB force field. Ligand parameters were generated with the General AMBER Force Field 2 (GAFF2) after restrained electrostatic potential (RESP) charges were derived using Gaussian 16. Each system was solvated in a TIP3P water box with a 10 Å buffer, and sufficient $Na^+$ and $Cl^-$ ions were added to neutralize the total charge.

To remove unfavorable contacts and achieve equilibration, a standard protocol was applied: two-stage energy minimization (first with positional restraints on the solute to relax the solvent, followed by unrestrained minimization of the whole system); heating from 0 K to 300 K over 50 ps under the constant-volume (NVT) ensemble; and 200 ps of equilibration under the constant-pressure (NPT) ensemble at 1.0 atm and 300 K. Production simulations were then performed for 100 ns under NPT conditions with all restraints removed. A 2 fs time step was used for a total of $5 \times 10^7$ steps, with trajectories and energies saved every 5,000 steps. To ensure statistical robustness, three independent runs per complex were performed with different random seeds.

Simulation trajectories were analyzed using the cpptraj module. Backbone root-mean-square deviation (RMSD) and per-residue root-mean-square fluctuation (RMSF) were computed. Binding free energies ($\Delta G_{binding}$) were estimated by the Molecular Mechanics Poisson–Boltzmann Surface Area (MM-PBSA) method using the final 10 ns of stable trajectories (1,000 frames) to provide a quantitative assessment of binding strength.

## 3. Results and Discussion

### 3.1 Performance Evaluation of Predictor Submodels

To ensure that the generated property spectrum accurately characterizes molecular bioactivity, the 33 independently constructed submodels were subjected to rigorous performance validation. All models were evaluated on independent test sets, and the results are summarized in Table 2.

Strong predictive performance was observed across both regression and classification tasks. For the 27 regression models (S1–S27), a mean coefficient of determination ($R^2$) of 0.90 was obtained. Models targeting physicochemical properties (S1–S11) exhibited the most robust performance, with a mean $R^2$ exceeding 0.91. Notably, high predictive accuracy was maintained for pharmacokinetic parameters that are typically difficult to model, including bioavailability (S12, $R^2 = 0.88$) and clearance (S16, $R^2 = 0.88$), indicating that deep neural networks have a clear advantage in capturing complex structure–activity relationships (SARs).

For the six classification models (S28–S33) related to safety and stability, the mean area under the receiver operating characteristic curve (ROC AUC) reached 0.94. In particular, Ames mutagenicity (S29) and human Ether-à-go-go-Related Gene (hERG) toxicity risk (S30) achieved AUC values of 0.96 and 0.95, respectively. These findings indicate that the ensemble of 33 submodels is sufficiently robust to provide a reliable basis for constructing the high-dimensional property spectrum and for computing the $S_{5F}$ similarity.

[Please insert Table 2 here.]

**3.2 Analysis of Property Spectra for Candidate Drugs**

On the basis of the trained submodels, 33-dimensional property spectra were constructed for the three reference drugs (Cefradine, Cefuroxime, and Ceftriaxone) and for the selected candidate molecules. To visualize the extent of global property alignment between candidates and references, the normalized property vectors were visualized as Property Fingerprint Spectra (Figure 5).

In these spectra, the x-axis denotes the 33 key properties, and the y-axis represents the normalized property scores (ranging from 0 to 1). The reference drugs (thick red line) display distinctive peak–trough patterns that constitute their bioactivity fingerprints; for example, a peak at the octanol–water partition coefficient (logP) indicates high lipophilicity, whereas a trough at toxicity risk indicates a favorable safety profile. For each reference drug, the top five candidates (thin blue lines) exhibit trajectories that are highly concordant with the corresponding reference. Under the imposed Euclidean distance criterion, alignment is achieved not only in overall trends but also in absolute magnitudes. This visual concordance provides strong evidence that the selected molecules do not merely perform well on isolated metrics but closely recapitulate the advantageous developability characteristics of the reference drugs across multiple dimensions, including physicochemical properties, pharmacokinetics, efficacy, and safety.

[Please insert Fig. 5 here.]

### 3.3 Virtual Screening Results and $S_{5F}$ Prioritization

On the basis of the composite $S_{5F}$ score (defined as a Euclidean distance–based similarity), the eMolecules, TargetMol, and MCE screening libraries were ranked globally. For each of the three reference drugs, the five most similar molecules were retained, yielding a total of fifteen lead candidates.

Detailed results are reported in Table 3. All selected molecules achieved $S_{5F}$ scores greater than 0.93, and the first-ranked molecule within each reference set consistently exceeded 0.955, indicating a very small mean Euclidean distance to the corresponding reference drug in the 33-dimensional property space. As an illustrative example, the top candidate associated with the Cefradine reference, M2, attained an $S_{5F}$ score of 0.956. Further analysis of the property spectra indicated that this molecule closely recapitulated the physicochemical and activity patterns of the reference; moreover, for the critical cardiac safety endpoint—human Ether-à-go-go-Related Gene (hERG) toxicity risk (S30)—its predicted probability was lower than that of the reference (0.55 versus 0.61), suggesting potentially superior safety.

Conventional virtual screening commonly relies on structural similarity computed from molecular fingerprints, such as Extended-Connectivity Fingerprints (ECFP4) with the Tanimoto coefficient. However, structural similarity does not strictly imply similarity in bioactivity and is vulnerable to "patent cliffs." In contrast, the property spectra developed in this study were aligned directly in phenotypic space. The results shown in Figure 5 indicate that, even when a candidate and its reference drug differ in chemical scaffold, overlapping 33-dimensional property spectra are associated with comparable developability potential. This strategy naturally supports scaffold hopping and facilitates the identification of novel molecules with new intellectual property (IP) while maintaining comparable efficacy.

Regarding the choice of similarity metric, the Euclidean distance–based $S_{5F}$, rather than cosine similarity, was adopted—a choice that is critical. Drug properties (for example, solubility and toxicity thresholds) possess strict absolute physical meaning. In safety assessment, for instance, toxicity probabilities of 0.1 (safe) and 0.9 (highly toxic) may align in vector direction yet diverge fundamentally in developability. As shown in Table 3, molecules with high $S_{5F}$ scores closely match the reference drugs in the absolute values of all key metrics, thereby avoiding false positives in which directional similarity masks magnitude offsets and ensuring the reliability of the screening outcomes.

Despite the strong model performance, certain limitations remain. The accuracy of the property spectra depends entirely on the predictive capability of the submodels. Although high coefficients of determination ($R^2$) were obtained on the test sets, predictive confidence may degrade for rare chemical scaffolds that are out of distribution (OOD) relative to the training data. Accordingly, subsequent experimental validation is essential.

[Please insert Table 3 here.]

### 3.4 Antibacterial Activity by Disk Diffusion

To rapidly identify compounds with measurable antibacterial activity from the fifteen MDL candidate molecules, a modified disk diffusion assay was first employed for phenotypic screening. As shown in Figure 6, the majority of candidates either failed to exhibit discernible inhibitory activity or were unable to diffuse effectively through the agar matrix owing to limited solubility. By contrast, M2, M8, and M9 produced well-defined zones of inhibition at a loading of 30 micrograms per disk, indicating the capacity to inhibit the growth of Escherichia coli ATCC 25922.

Specifically, M2 exhibited the strongest activity, with a mean inhibition zone diameter of 20.26 mm, approaching that of the positive control drug, Cefuroxime (22.03 mm). In comparison, although a very high score was obtained for M8 in subsequent molecular docking, its inhibition zone measured only 14.46 mm, markedly smaller than that of M2; this outcome is plausibly attributable to suboptimal physicochemical properties that limited diffusion in the aqueous agar environment. M9 showed intermediate activity (17.65 mm). On the basis of these phenotypic prescreening results, M2, M8, and M9 were selected as target compounds for subsequent quantitative determination of the minimum inhibitory concentration (MIC) and the minimum bactericidal concentration (MBC).

[Please insert Fig. 6 here.]

## 3.5 MIC and MBC Determination Results

In accordance with the Clinical and Laboratory Standards Institute (CLSI) M07-A9 guideline, the broth microdilution assay was used to quantify the in vitro antibacterial activity of the three selected molecules (M2, M8, and M9) and the reference antibiotic, cefuroxime. As shown in Table 4, the minimum inhibitory concentrations (MICs) against Escherichia coli ATCC 25922 were 25.6 ug/mL for M2 and M8, compared with 51.2 ug/mL for M9.

Although the candidates exhibited MIC values higher than that of the approved drug cefuroxime (3.2 ug/mL), clear inhibitory activity was already observed for M2 at 25.6 ug/mL as an unoptimized hit compound, indicating considerable development potential. With respect to bactericidal potency, the minimum bactericidal concentration (MBC) of M2 was 51.2 ug/mL, substantially lower than those of M8 and M9 (both 102.4 ug/mL). The resulting MBC/MIC ratios for M2, M8, and M9 were 2, 4, and 2, respectively, all meeting the conventional criterion of an MBC/MIC ratio less than or equal to 4, indicating bactericidal rather than merely bacteriostatic activity against Escherichia coli ATCC 25922.

Considering both MIC and MBC, M2 was identified as the lead compound with the greatest developability potential in this study, as it maintained a comparatively low MIC while exhibiting stronger bactericidal activity (lower MBC).

[Please insert Table 4 here.]

**3.6 Molecular Docking Results**

Molecular docking was carried out for the top fifteen candidate molecules and three marketed reference antibiotics to assess their potential affinity for penicillin-binding protein 2 (PBP2). As summarized in Table 5, nearly all tested molecules (three references and fifteen candidates) achieved their most favorable (lowest) docking scores within Cavity 1, a pocket with an estimated volume of approximately 517 $Å^3$. This finding suggests that Cavity 1 is likely the active site or primary binding pocket of PBP2, whose spatial architecture and physicochemical environment provide optimal binding and orientation for cephalosporin-class ligands, thereby facilitating acylation and leading to covalent inhibition of the protein.

The mean docking score of the top fifteen candidates (-9.9 kcal/mol) was clearly superior to that of the three marketed drugs (mean -8.4 kcal/mol), supporting the ability of the deep learning model to identify high-affinity scaffolds. Upon further analysis, M8 achieved the most favorable docking score (-11.3 kcal/mol), markedly surpassing cefuroxime (-8.5 kcal/mol). However, this high theoretical affinity did not translate into superior in vitro antibacterial activity, as M8 yielded the smallest zone of inhibition, indicating potential developability liabilities such as limited solubility or membrane permeability. By contrast, M2 yielded a docking score of -9.8 kcal/mol, slightly less favorable than M8 yet still substantially better than the positive control. Taken together with its leading performance in the disk diffusion and minimum inhibitory concentration (MIC) assays, M2 was considered to strike a more favorable balance between target-binding affinity and physicochemical properties and/or membrane permeability.

[Please insert Table 5 here.]

Structural analysis was conducted and revealed molecular-level differences in how the compounds engaged the active pocket of penicillin-binding protein 2 (PBP2), providing an atomistic basis for the distribution of in vitro antibacterial activity and the observed structure-activity relationships (Figure 7). As the positive control exhibiting the strongest in vitro activity, cefuroxime was docked in a pre-reaction (pre-acylation) state. Its binding was stabilized primarily by pi-alkyl hydrophobic interactions between ALA65 and the beta-lactam ring, which drove the core scaffold into the active groove, and was complemented by a conventional hydrogen bond between ARG68 and the carbamate side chain that acted as a directional anchor to lock the pose. Although its noncovalent binding energy (-8.5 kcal/mol) was not the most favorable, this specific orientation is a prerequisite for subsequent nucleophilic attack by the catalytic serine and formation of a covalent adduct, thereby explaining its exceptional activity as a covalent inhibitor.

Among the selected candidates, the best-performing molecule, M2 (MIC = 25.6 µg/mL), was observed to employ a pronounced "polar locking" mechanism that compensates for limitations in noncovalent binding energy. As shown in Figure 7B, two strong conventional hydrogen bonds were formed between the sultam ring and sulfonamide side chain of M2 and residues THR202 and LYS162, establishing a dual polar core for binding. In addition, an extensive network of carbon-hydrogen bonds with LYS159 and ALA201 further restricted conformational freedom. This combination of strong hydrogen-bond anchoring, auxiliary weak hydrogen bonds, and hydrophobic support at ALA65 substantially enhanced binding specificity, resulting in antibacterial potency approaching that of a covalent agent.

Although M8 achieved the most favorable theoretical docking score (-11.3 kcal/mol), its binding mode appeared to rely excessively on electrostatic and hydrophobic interactions. The nitrobenzofuranone scaffold of M8 was embedded in a strong electrostatic field created by GLU89/GLU157 (pi-anion) and LYS159 (pi-cation), while an extensive hydrophobic network involving VAL179 and ARG68 was simultaneously engaged. This "high-energy adhesion" elevated the computed score; however, excessive hydrophobicity combined with a complex charge environment may impair solubility and diffusion behavior under physiological conditions. This interpretation is consistent with its relatively small zone of inhibition in the disk diffusion assay (14.46 mm), indicating potential liabilities in developability.

M9 exemplified a distinct binding mode. A strongly polar sulfonic acid group on the right-hand portion of the molecule formed key hydrogen bonds with THR202 and PRO66, conferring favorable aqueous solubility and accounting for its comparatively large inhibition zone. Nevertheless, a critical liability was identified: an unfavorable donor-donor interaction between a nitrogen atom within the core scaffold and the key residue LYS162 (Figure 7D, red dashed line). This local repulsion imposes an energetic penalty on binding, explaining why M9, despite engaging the pocket, exhibited a substantially higher minimum inhibitory concentration (51.2 μg/mL) than M2. This finding also indicates a clear optimization direction for the M9 scaffold, namely, fine-tuning to eliminate the steric and/or electrostatic clash.

[Please insert Fig. 7 here.]

### 3.7 Molecular Dynamics Simulation

To further substantiate the docking results, molecular dynamics simulations were carried out on the penicillin-binding protein 2 (PBP2)-ligand complexes. Root-mean-square deviation (RMSD) and root-mean-square fluctuation (RMSF) were analyzed, and binding free energies were evaluated using the molecular mechanics Poisson-Boltzmann surface area (MM-PBSA) method.

First, overall structural stability and residue-level flexibility were assessed by RMSD and RMSF, respectively. The initial structures were used as references, and the time evolution of RMSD for both systems is shown in Figure 8A. Within the first 20 nanoseconds, all trajectories equilibrated rapidly, supporting the validity of the initial docking poses. After equilibration, the M2 complex (red/orange/green traces) exhibited marked conformational convergence, with a mean RMSD of 2.8 $\pm$ 0.2 Å, notably lower than that of the positive control cefuroxime (black/purple/cyan traces) at 3.3 $\pm$

0.3 Å. This approximately 0.5 Å difference is mechanistically informative: it suggests that cefuroxime, as a covalent inhibitor, occupies a relatively loose noncovalent pre-reaction state that retains substantial conformational freedom prior to covalent bond formation, whereas M2, despite lacking the "permanent locking" conferred by a covalent bond, induces global rigidification of PBP2 already at the physical binding stage via a dense hydrogen-bond network formed by its distinctive sultam scaffold within the active pocket. Such physical stability, exceeding that of the covalent drug's pre-reaction state, provides a compelling explanation for M2's potent antibacterial activity as a noncovalent inhibitor.

Subsequent RMSF analysis (Figure 8B) localized these stability differences to functionally critical regions of PBP2 and revealed a "dynamic locking" mechanism at the active site. Although the overall fluctuation profiles of M2 (red) and cefuroxime (black) were similar, a pronounced divergence was observed for residues 50-100: cefuroxime displayed a prominent flexibility peak (approximately 4.5 Å), whereas M2 suppressed atomic fluctuations in this region to approximately 2.0 Å. Given that this segment includes the core hydrophobic anchor ALA65, the result indicates that the pi-alkyl interaction between M2 and ALA65 remains exceptionally robust under dynamic conditions, effectively constraining the conformational breathing of this key loop. Moreover, in the catalytic center (residues 190-210), which contains critical hydrogen-bonding sites THR202 and ALA201, M2 similarly induced lower RMSF values, indicating that its side-chain polar contacts act as "molecular rivets." Collectively, by selectively damping the dynamics of the active pocket and key loops, M2 was shown to impede substrate entry and catalysis, thereby elucidating at the atomic scale the molecular basis of its strong bactericidal activity.

[Please insert Fig. 8 here.]

To quantify interaction strength, binding free energies for the final 10 ns of each trajectory were calculated using the molecular mechanics Poisson-Boltzmann surface area (MM-PBSA) method. A binding free energy of -38.54 kcal/mol was obtained for the penicillin-binding protein 2 (PBP2)-M2 complex, which was markedly more favorable than that of the cefuroxime-PBP2 complex (-29.40 kcal/mol).

From energy decomposition, binding of M2 was found to be driven predominantly by van der Waals (-45.21 kcal/mol) and electrostatic (-28.65 kcal/mol) terms. Although a sizable unfavorable polar solvation contribution was present (+39.12 kcal/mol), the net binding free energy remained superior. By contrast, weaker van der Waals (-35.40 kcal/mol) and electrostatic (-18.22 kcal/mol) contributions were obtained for cefuroxime. Taken together, these energetic analyses quantitatively corroborate the superior noncovalent binding of M2.

## 4. Conclusions

A screening strategy termed five-fold maximum drug-likeness (5F-MDL) was introduced and validated to address the global challenge of discovering new agents against multidrug-resistant Escherichia coli (MDR E. coli). In contrast to traditional paradigms based on substructure similarity, a fundamental shift toward high-dimensional phenotypic property alignment was realized. An evaluation system comprising 33 well-calibrated deep learning submodels (regression coefficient of determination $R^2 \approx 0.90$; classification area under the receiver operating characteristic curve (AUC) $\approx 0.94$) was constructed to translate the abstract notion of drug-likeness into a quantifiable property spectrum. A five-fold score ($S_{5F}$) was introduced as the core metric, by which scaffold-hopping chemotypes closely mirroring clinically used drugs in physicochemical, pharmacokinetic, and safety attributes were identified across the a library of 16-million molecules.

The translational efficiency of the strategy was substantiated by biological validation. Among 15 prioritized candidates, M2, M8, and M9 demonstrated clear in vitro antibacterial activity. Notably, the lead compound M2, owing to a favorable balance of properties, produced a zone of inhibition in disk diffusion assays second only to the positive control, and achieved a minimum inhibitory concentration (MIC) of 25.6 μg/mL together with a minimum bactericidal concentration (MBC) to MIC ratio of 2 or less, thereby confirming a bactericidal profile. Comparative analyses further underscored the distinctive advantage of 5F-MDL in excluding "high-score, low-performance" molecules: although M8 achieved the highest theoretical docking score, suboptimal physicochemical properties limited its realized activity, whereas the efficacy of M9 was constrained by local steric clashes between its core scaffold and the receptor.

Mechanistic studies elucidated the mode of action of M2 as a noncovalent inhibitor of penicillin-binding protein 2 (PBP2). Molecular docking and molecular dynamics simulations indicated that, unlike cefuroxime, which relies on covalent chemistry, M2 adopts a distinct binding mode characterized by hydrophobic anchoring together with polar locking. Hydrophobic interactions with ALA65 were retained, while a dense hydrogen-bond network mediated by the sultam ring and the side chain with THR202 and LYS162 secured tight engagement with the active pocket. In 100 ns molecular dynamics simulations, the backbone root-mean-square deviation (RMSD) of the M2–PBP2 complex (2.8 Å) was significantly lower than that of the noncovalent pre-reaction state of cefuroxime (3.3 Å), and selective "dynamic freezing" of the active-site loop containing ALA65 was induced by M2. These observations indicate that purely physical binding by M2 can achieve, and even exceed, the conformational stability associated with the covalent pre-reaction state.

In summary, a noncovalent PBP2 inhibitor with promising clinical development potential (M2) was identified, and the scientific validity of a screening logic based on panoramic alignment of properties with clinically approved drugs was substantiated. By subjecting candidates to a rigorous 33-dimensional assessment of drug-likeness, the 5F-MDL framework ensures that selected compounds not only meet activity thresholds but also display characteristics associated with mature therapeutics. This strategy provides a methodological tool for overcoming the high attrition that hampers antimicrobial drug discovery against resistant pathogens and is expected to accelerate progression from hits to clinical candidates.

**Acknowledgements**

This work was supported by the National Natural Science Foundation of China (grant numbers 82260896).

**Captions of figures and tables:**

**Fig. 1** The central concept of five-fold maximum drug-likeness (5F-MDL).

**Fig. 2** Chemical structures of three reference inhibitors of penicillin-binding protein 2 (PBP2) (reference drugs).

**Fig. 3** Overall architecture of the five-fold maximum drug-likeness (5F-MDL) computational framework.

**Fig. 4** Canonical network architecture of an individual deep learning submodel for property prediction.

**Fig. 5** Panoramic comparison of profiles in the 33-dimensional property space for the three reference drugs—(a) cefradine; (b) cefuroxime; (c) ceftriaxone—and their respective top five candidate molecules. The x-axis is indexed by the 33 properties, and the y-axis displays normalized predicted values. The reference drug is shown as a thick red line, and the corresponding top five candidates are shown as thin blue lines. Extensive overlap between curves indicates very high similarity in the multidimensional property space.

**Fig. 6** In vitro antibacterial activity against Escherichia coli ATCC 25922 of the prioritized candidates (M2, M8, M9) and the reference antibiotic, cefuroxime.

**Fig. 7** Binding modes predicted by molecular docking of the reference drugs (a) cefuroxime and the three best candidate molecules (b) M2; (c) M8; and (d) M9 in the PBP2 target.

**Fig. 8** Stability analysis from molecular dynamics simulations of complexes formed by cefuroxime or candidate M2 with PBP2 (a) time trajectories of the backbone root-mean-

square deviation (RMSD) of the complexes; and (b) root-mean-square fluctuation (RMSF) profiles of protein residues.

**Table 1** Composition of the 33 property datasets and an overview of the corresponding predictive submodels.

**Table 2** Predictive performance of the 33 property-specific submodels on independent test sets.

**Table 3** Overview of the top fifteen candidates with high drug-likeness, their corresponding reference drugs, and similarity scores.

**Table 4** Minimum inhibitory concentration (MIC) and minimum bactericidal concentration (MBC) of the prioritized active molecules and the reference antibiotics.

**Table 5** Comparison of molecular docking scores for the three reference drugs and the top fifteen candidate molecules.

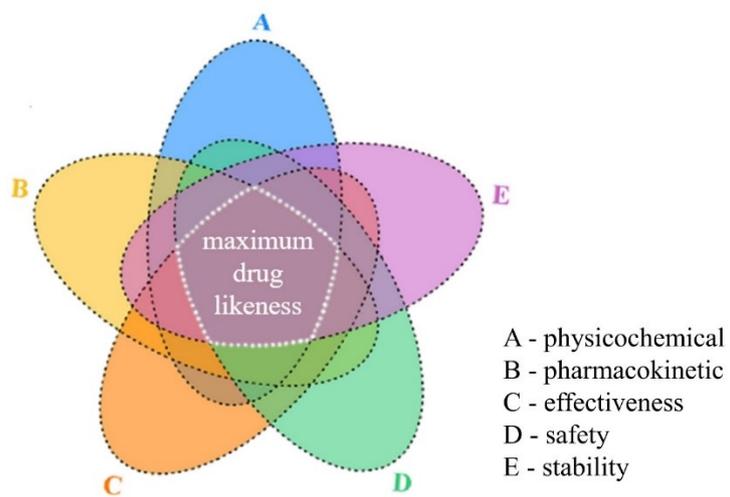

**Figure 1** The central concept of fivefold maximum drug-likeness (5F-MDL).

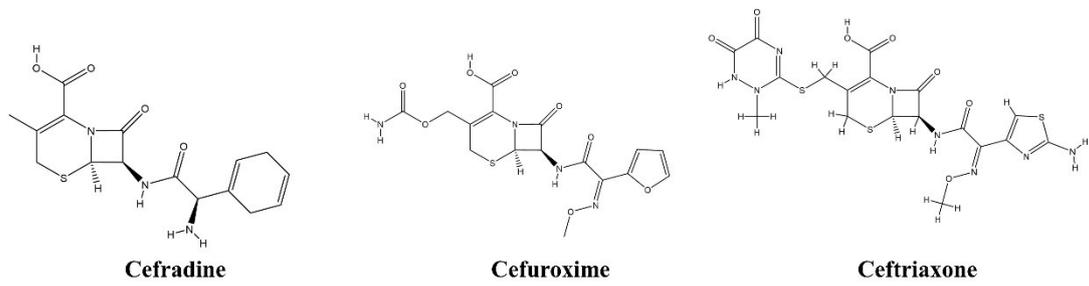

**Figure 2** Chemical structures of three reference inhibitors of penicillin-binding protein 2 (PBP2) (reference drugs).

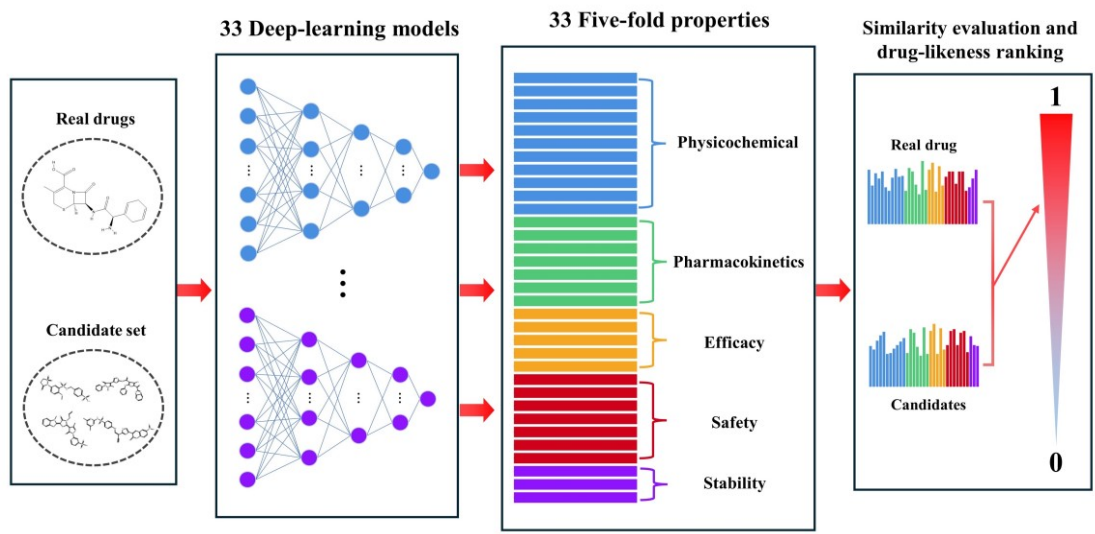

**Figure 3** Overall architecture of the five-fold maximum drug-likeness (5F-MDL) for drug screening.

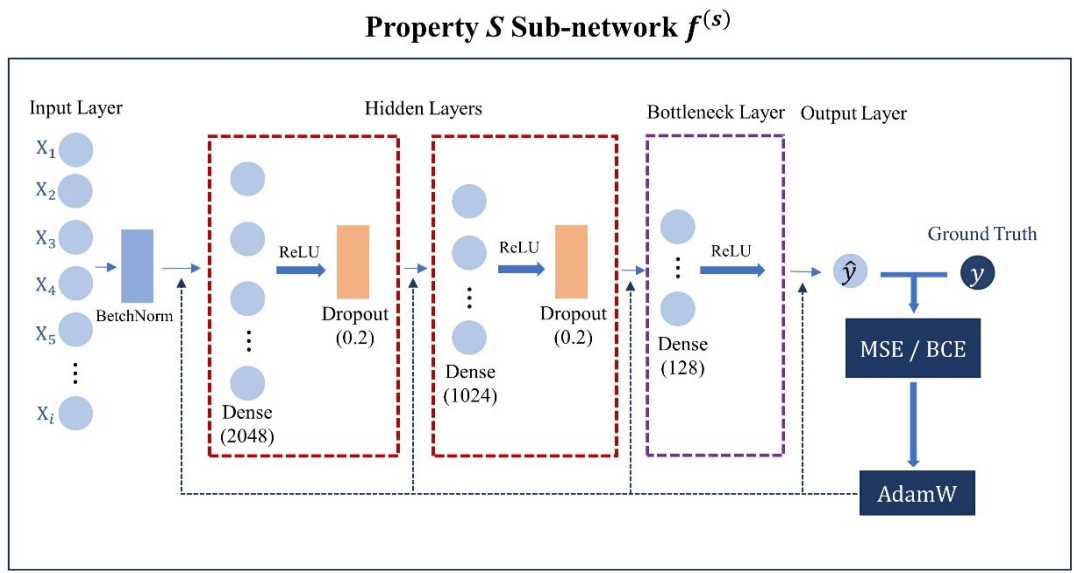

**Figure 4** Canonical network architecture of an individual deep learning submodel for property prediction.

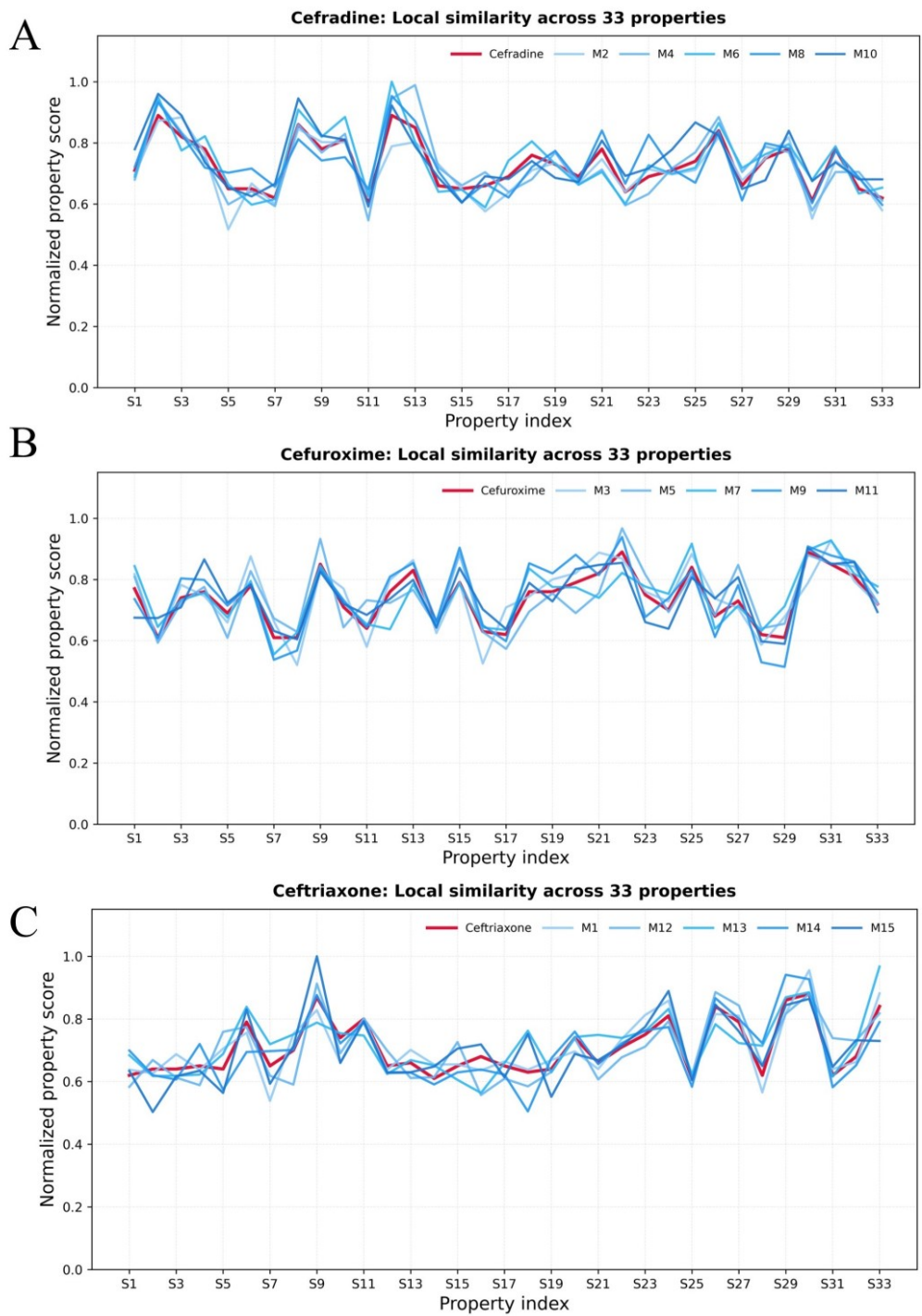

**Figure 5** Panoramic comparison of profiles in the 33-dimensional property space for the three reference drugs—(a) cefradine; (b) cefuroxime; (c) ceftriaxone—and their respective top five candidate molecules. The x-axis is indexed by the 33 properties, and the y-axis displays normalized predicted values. The reference drug is shown as a thick red line, and the corresponding top five candidates are shown as thin blue lines. Extensive overlap between curves indicates very high similarity in the multidimensional property space.

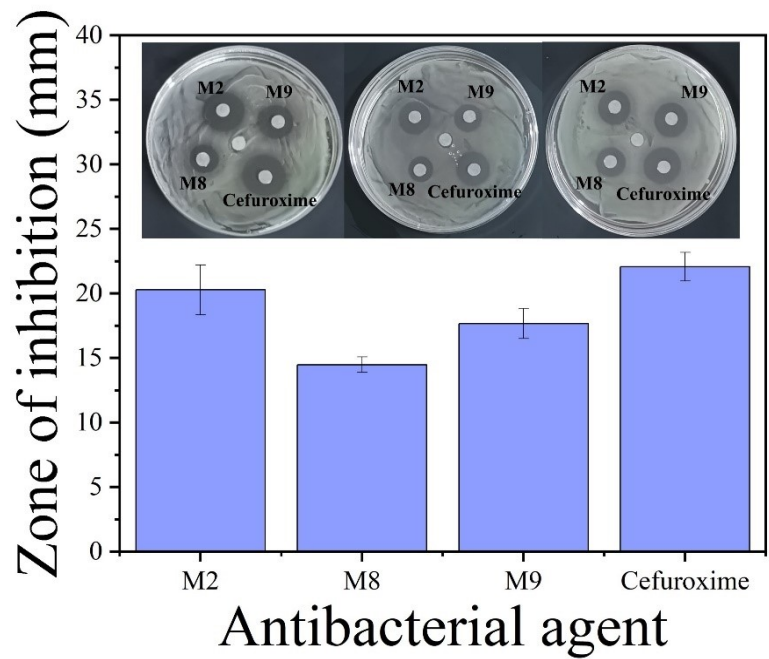

**Figure 6** In vitro antibacterial activity against Escherichia coli ATCC 25922 of the prioritized candidates (M2, M8, M9) and the reference antibiotic, cefuroxime.

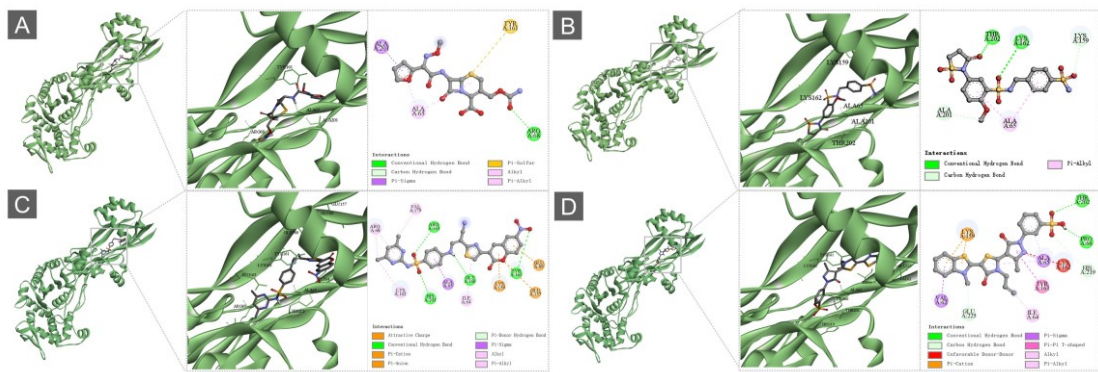

**Figure 7** Binding modes predicted by molecular docking of the reference drugs (a) cefuroxime and the three best candidate molecules (b) M2; (c) M8; and (d) M9 in the PBP2 target.

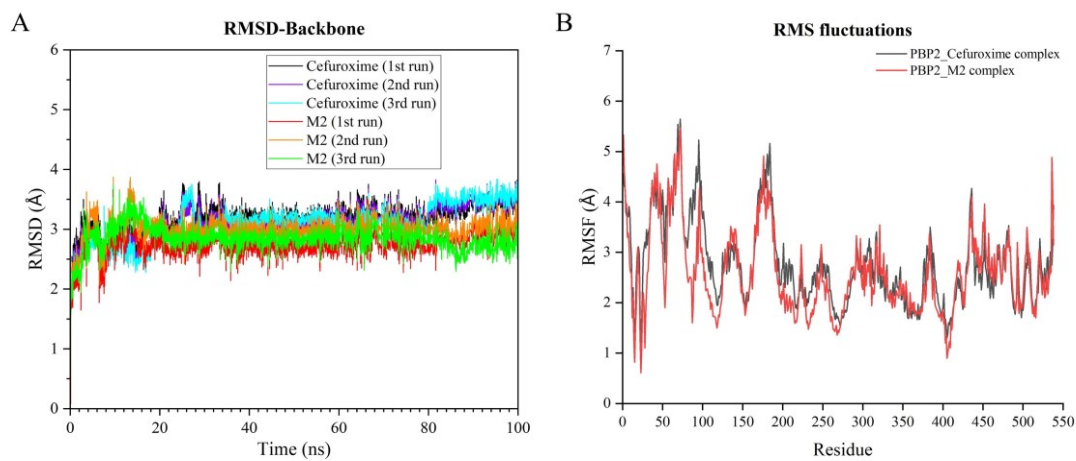

**Figure 8** Stability analysis from molecular dynamics simulations of complexes formed by cefuroxime or candidate M2 with PBP2 (a) time trajectories of the backbone root-mean-square deviation (RMSD) of the complexes; and (b) root-mean-square fluctuation (RMSF) profiles of protein residues.

**Table 1** Composition of the 33 property datasets and an overview of the corresponding predictive submodels.

| No. | Name | Class | Training set | Validation set | Testing set |
|---|---|---|---|---|---|
| S1 | Aqueous solubility (LogS) | C1 | 251267 | 53847 | 53851 |
| S2 | Octanol–water partition coefficient (log P) | C1 | 286544 | 61254 | 61263 |
| S3 | Melting point | C1 | 130559 | 27972 | 27984 |
| S4 | Boiling point | C1 | 125784 | 26958 | 26962 |
| S5 | Surface tension | C1 | 59168 | 12679 | 12697 |
| S6 | Density | C1 | 167325 | 35846 | 35869 |
| S7 | Viscosity | C1 | 109081 | 23374 | 23392 |
| S8 | Flash point | C1 | 161846 | 34675 | 34683 |
| S9 | Vapor pressure | C1 | 137928 | 29558 | 29564 |
| S10 | Dissociation constant | C1 | 257753 | 55163 | 55319 |
| S11 | Hydrolysis (half-life value) | C1 | 117046 | 25098 | 25183 |
| S12 | Bioavailability | C2 | 39298 | 8417 | 8461 |
| S13 | Plasma protein binding rate | C2 | 77264 | 16547 | 16583 |
| S14 | Maximal rate of metabolism | C2 | 102937 | 22054 | 22068 |
| S15 | Biliary excretion rate | C2 | 44518 | 9539 | 9671 |
| S16 | Urinary excretion rate | C2 | 57047 | 12248 | 12239 |
| S17 | Volume of distribution | C2 | 137489 | 29439 | 29586 |
| S18 | Half-life | C2 | 129632 | 27747 | 27864 |
| S19 | Minimum Inhibitory Concentration (MIC) | C3 | 335587 | 71956 | 71998 |
| S20 | Enzyme inhibition constant (Ki) | C3 | 137418 | 29447 | 29456 |
| S21 | Receptor affinity | C3 | 185813 | 39736 | 39973 |
| S22 | Maximum effect model parameter (Emax) | C3 | 48246 | 10348 | 10369 |
| S23 | 50% effective dose (EC50) | C3 | 161724 | 34648 | 34623 |
| S24 | Median lethal dose (LD50) | C4 | 161591 | 34627 | 34654 |
| S25 | No Observed Adverse Effect Level (NOAEL) | C4 | 206658 | 44246 | 44289 |
| S26 | Tetrahymena pyriformis 50% growth inhibition concentration | C4 | 245736 | 52639 | 52784 |
| S27 | Median lethal concentration (LC50) | C4 | 155369 | 33247 | 33286 |
| S28 | Developmental toxicity | C4 | 257561 | 55142 | 55364 |
| S29 | Ames mutagenicity | C4 | 397734 | 85237 | 85368 |
| S30 | hERG_risk | C4 | 174397 | 37296 | 37412 |
| S31 | Property Name | C5 | 249362 | 53447 | 53586 |
| S32 | Chemical stability | C5 | 335341 | 71839 | 71964 |
| S33 | Thermostability | C5 | 33789 | 7228 | 7243 |

*C1, physicochemical properties; C2, pharmacokinetics; C3, efficacy; C4, safety; C5, stability.

**Table 2** Predictive performance of the 33 property-specific submodels on independent test sets.

| Number | Task Type | Evaluation Metric | Performance |
|---|---|---|---|
| S1 | Regression | $R^2$ | 0.87 |
| S2 | Regression | $R^2$ | 0.89 |
| S3 | Regression | $R^2$ | 0.86 |
| S4 | Regression | $R^2$ | 0.92 |
| S5 | Regression | $R^2$ | 0.94 |
| S6 | Regression | $R^2$ | 0.89 |
| S7 | Regression | $R^2$ | 0.88 |
| S8 | Regression | $R^2$ | 0.91 |
| S9 | Regression | $R^2$ | 0.93 |
| S10 | Regression | $R^2$ | 0.95 |
| S11 | Regression | $R^2$ | 0.92 |
| S12 | Regression | $R^2$ | 0.88 |
| S13 | Regression | $R^2$ | 0.86 |
| S14 | Regression | $R^2$ | 0.87 |
| S15 | Regression | $R^2$ | 0.86 |
| S16 | Regression | $R^2$ | 0.88 |
| S17 | Regression | $R^2$ | 0.89 |
| S18 | Regression | $R^2$ | 0.89 |
| S19 | Regression | $R^2$ | 0.93 |
| S20 | Regression | $R^2$ | 0.95 |
| S21 | Regression | $R^2$ | 0.91 |
| S22 | Regression | $R^2$ | 0.93 |
| S23 | Regression | $R^2$ | 0.92 |
| S24 | Regression | $R^2$ | 0.93 |
| S25 | Regression | $R^2$ | 0.91 |
| S26 | Regression | $R^2$ | 0.92 |
| S27 | Regression | $R^2$ | 0.96 |
| S28 | Classification | AUC | 0.93 |
| S29 | Classification | AUC | 0.96 |
| S30 | Classification | AUC | 0.95 |
| S31 | Classification | AUC | 0.96 |
| S32 | Classification | AUC | 0.91 |
| S33 | Classification | AUC | 0.92 |

**Table 3** Overview of the top fifteen candidates with high drug-likeness, their corresponding reference drugs, and similarity scores.

| No. | Specs ID | Structure | $S_{5F}$ | Nearest drug | Euclidean Dist. |
|---|---|---|---|---|---|
| M1 | AS-871/40992156 | 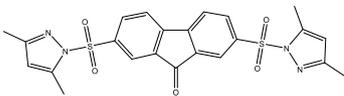 | 0.956 | Ceftriaxone | 0.252 |
| M2 | AS-871/43476343 | 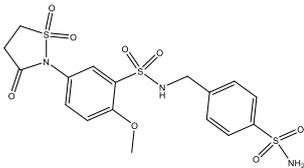 | 0.956 | Cefuroxime | 0.253 |
| M3 | AJ-916/13003001 | 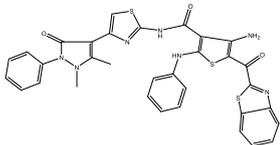 | 0.956 | Cefradine | 0.254 |
| M4 | AE-641/10060008 | 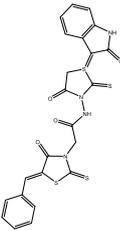 | 0.952 | Cefradine | 0.274 |
| M5 | AG-205/07689035 | 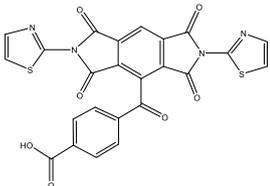 | 0.951 | Cefuroxime | 0.281 |
| M6 | AG-690/08553010 | 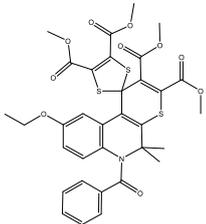 | 0.950 | Cefradine | 0.286 |
| M7 | AG-690/08992014 | 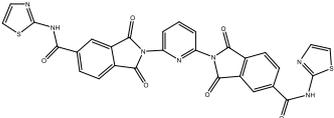 | 0.950 | Cefuroxime | 0.287 |
| M8 | AE-848/11488873 | 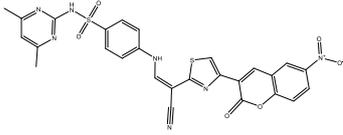 | 0.950 | Cefradine | 0.290 |
| M9 | AG-690/12867888 | 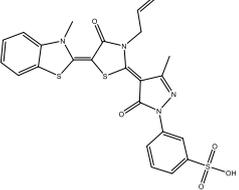 | 0.949 | Cefuroxime | 0.292 |
| M10 | AG-690/12249326 | 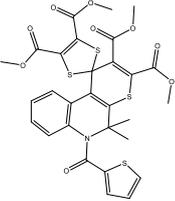 | 0.946 | Cefradine | 0.313 |
| M11 | AS-871/40777464 | 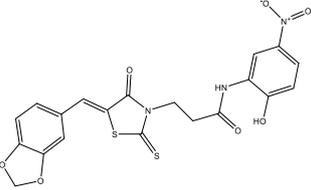 | 0.945 | Cefuroxime | 0.318 |
| M12 | AN-989/14323005 | 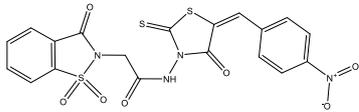 | 0.944 | Ceftriaxone | 0.320 |
| M13 | AO-079/14332007 | 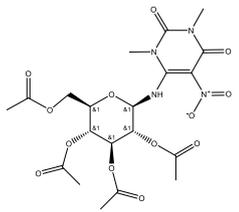 | 0.942 | Ceftriaxone | 0.334 |
| M14 | AG-205/14365087 | 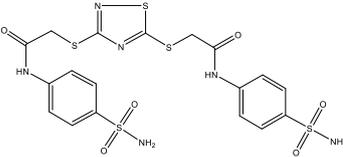 | 0.941 | Ceftriaxone | 0.337 |
| M15 | AB-323/13887102 | 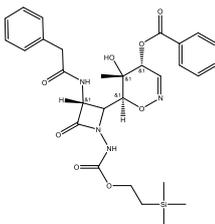 | 0.936 | Ceftriaxone | 0.370 |

**Table 4** Minimum inhibitory concentration (MIC) and minimum bactericidal concentration (MBC) of the prioritized active molecules and the reference antibiotics.

| Compd. no | MIC | MBC |
|---|---|---|
| M2 | 25.6 | 51.2 |
| M8 | 25.6 | 102.4 |
| M9 | 51.2 | 102.4 |
| Cefuroxime | 3.2 | 6.4 |

**Table 5** Comparison of molecular docking scores for the three reference drugs and the top fifteen candidate molecules.

| Cavity Volumes | | | | | | | | | |
|---|---|---|---|---|---|---|---|---|---|
| 517 Å³ | | 549 Å³ | | 286 Å³ | | 635 Å³ | | 510 Å³ | |
| Compd. no | kcal/mol | Compd. no | kcal/mol | Compd. no | kcal/mol | Compd. no | kcal/mol | Compd. no | kcal/mol |
| Cefradine | -8.2 | Cefradine | -7.6 | Cefradine | -7.3 | Cefradine | -6.5 | Cefradine | -6.1 |
| Cefuroxime | -8.5 | Cefuroxime | -7.6 | Cefuroxime | -6.5 | Cefuroxime | -6.1 | Cefuroxime | -6.8 |
| Ceftriaxone | -8.4 | Ceftriaxone | -8.2 | Ceftriaxone | -7.5 | Ceftriaxone | -7.2 | Ceftriaxone | -7.9 |
| M1 | -10.9 | M1 | -10.0 | M1 | -9.2 | M1 | -8.7 | M1 | -8.2 |
| M2 | -9.8 | M2 | -7.7 | M2 | -7.8 | M2 | -7.7 | M2 | -7.0 |
| M3 | -11.2 | M3 | -9.9 | M3 | -9.5 | M3 | -8.5 | M3 | -9.2 |
| M4 | -9.0 | M4 | -8.8 | M4 | -7.3 | M4 | -7.2 | M4 | -7.8 |
| M5 | -9.1 | M5 | -8.9 | M5 | -8.2 | M5 | -7.5 | M5 | -7.8 |
| M6 | -10.2 | M6 | -7.2 | M6 | -5.5 | M6 | -5.4 | M6 | -6.4 |
| M7 | -11.0 | M7 | -10.2 | M7 | -10.6 | M7 | -8.8 | M7 | -9.4 |
| M8 | -11.3 | M8 | -10.0 | M8 | -11.2 | M8 | -9.0 | M8 | -8.7 |
| M9 | -10.5 | M9 | -8.7 | M9 | -8.9 | M9 | -7.4 | M9 | -7.8 |
| M10 | -8.7 | M10 | -10.2 | M10 | -7.1 | M10 | -6.9 | M10 | -6.2 |
| M11 | -9.5 | M11 | -8.5 | M11 | -8.5 | M11 | -7.4 | M11 | -7.3 |
| M12 | -9.1 | M12 | -9.0 | M12 | -8.5 | M12 | -7.8 | M12 | -7.8 |
| M13 | -9.0 | M13 | -8.0 | M13 | -7.8 | M13 | -6.4 | M13 | -6.4 |
| M14 | -9.3 | M14 | -8.3 | M14 | -8.6 | M14 | -7.1 | M14 | -7.1 |
| M15 | -9.0 | M15 | -7.2 | M15 | -6.6 | M15 | -6.2 | M15 | -6.2 |